\documentclass[sigconf]{acmart}

\usepackage{enumitem}
\usepackage{booktabs}
\usepackage{subcaption}
\usepackage{graphicx}

\newdimen\figrasterwd
\figrasterwd\textwidth

\copyrightyear{2022} 
\acmYear{2022} 
\setcopyright{acmlicensed}\acmConference[CHI '22]{CHI Conference on Human Factors in Computing Systems}{April 29-May 5, 2022}{New Orleans, LA, USA}
\acmBooktitle{CHI Conference on Human Factors in Computing Systems (CHI '22), April 29-May 5, 2022, New Orleans, LA, USA}
\acmPrice{15.00}
\acmDOI{10.1145/3491102.3517511}
\acmISBN{978-1-4503-9157-3/22/04}

\begin{document}

\title[GANSpiration]{GANSpiration: Balancing Targeted and Serendipitous Inspiration in User Interface Design with Style-Based Generative Adversarial Network}

\author{Mohammad Amin Mozaffari}
\email{mohammad-amin.mozaffari@polymtl.ca}
\affiliation{%
  \institution{Polytechnique Montreal}
  \city{Montreal}
  \state{QC}
  \country{Canada}
}

\author{Xinyuan Zhang}
\email{xinyuan.zhang@polymtl.ca}
\affiliation{%
  \institution{Polytechnique Montreal}
  \city{Montreal}
  \state{QC}
  \country{Canada}
}

\author{Jinghui Cheng}
\email{jinghui.cheng@polymtl.ca}
\affiliation{%
  \institution{Polytechnique Montreal}
  \city{Montreal}
  \state{QC}
  \country{Canada}
}

\author{Jin L.C. Guo}
\email{jguo@cs.mcgill.ca}
\affiliation{%
  \institution{McGill University}
  \city{Montreal}
  \state{QC}
  \country{Canada}
}

%Changes:
% - Addressed the additional comments raised by reviewer 2 by clarifying the baseline and the perceptual model
% - Misc editing and formatting fixes

\begin{abstract}
Inspiration from design examples plays a crucial role in the creative process of user interface design. However, current tools and techniques that support inspiration usually only focus on example browsing with limited user control or similarity-based example retrieval, leading to undesirable design outcomes such as focus drift and design fixation. To address these issues, we propose the GANSpiration approach that suggests design examples for both targeted and serendipitous inspiration, leveraging a style-based Generative Adversarial Network. A quantitative evaluation revealed that the outputs of GANSpiration-based example suggestion approaches are relevant to the input design, and at the same time include diverse instances. A user study with professional UI/UX practitioners showed that the examples suggested by our approach serve as viable sources of inspiration for overall design concepts and specific design elements. Overall, our work paves the road of using advanced generative machine learning techniques in supporting the creative design practice.
\end{abstract}

% ACM Classfication
% Please use the 2012 Classifiers and see this link to embed them in the text: \url{https://dl.acm.org/ccs/ccs_flat.cfm}


% Author Keywords

\maketitle
% Run: latexdiff --flatten ./v1/main.tex ./v2/main.tex > diff.tex
% Response: https://docs.google.com/document/d/1NBbhnoB4B3jhROmjA2bEUVAfHd-LecNBDqzneQk6KMM

\section{Introduction}
User interface designers take on daily challenges in creating effective, usable, and innovative design work. Inspiration from existing design examples plays an essential role in this creative process~\cite{Goncalves2014}. Previous studies have observed that designers often actively seek, combine, and transform design examples to draw an analogy from existing full-fledged designs embedded with similar yet provocative ideas~\cite{Herring2009CHI}.

Current tools that support this inspirational activity usually fall into one of the two categories. First, the proliferating design gallery platforms, such as Dribbble~\cite{dribbble} and Behance~\cite{behance} support a bottom-up, serendipitous inspirational process where the designer examines a collection of designs, seemingly without a goal, in order to find ``interesting'' examples helpful to their work. Second, some recent work proposed design inspirational tools focused on suggesting examples based on certain types of design input (e.g. in the form of a sketch or an existing design), usually leveraging algorithms to determine image similarity~\cite{Swearngin2018, Ritchie2011, Lee2010, Behrang2018}; this represents a top-down, targeted process where the designer has a concrete idea in mind and seeks examples that implemented the idea.

While providing important inspirational support, both types of approaches have limitations. Our preliminary interaction with designers reveals that seeking inspiration from the design galleries can sometimes be an overwhelming experience and result in ``design drift'' (i.e. shifted design ideas from the original focus). On the other hand, over-exposure to examples with similar styles might cause design fixation (i.e. ``a blind adherence to a set of ideas or concepts''~\cite{Jansson1991}) that hinders the novelty of the design work~\cite{Marsh1996,Herring2009CHI}.

To address the limitations of existing design supporting tools, we attempt to seek a balance between targeted and serendipitous inspirations in this work. Particularly, we propose \textit{GANSpiration}, a set of approaches that use a style-based generative adversarial network (StyleGAN), trained with a large dataset of existing interface designs, to generate a diverse and yet focused set of examples based on a preliminary design input. StyleGAN is built on the Generative adversarial network (GAN), a machine learning framework that is comprised of two neural networks trained jointly (i.e. a generative network and a discriminative network)~\cite{Goodfellow2014GAN}. The generative network in StyleGAN, in particular, takes ``style inputs'' at different stages during the process of image synthesis, controlling the style of the generated images at different levels of granularity and details~\cite{karras2018stylebased}.

GANSpiration-based approaches leverage the StyleGAN technique to perform style transfer and generate new design artifacts based on existing design artifacts and, therefore, provide a targeted and serendipitous inspiration for user interface designers. These approaches take a preliminary user interface design as input, merge the input image with a random set of existing designs using StyleGAN, and output representative examples that are either directly synthesized or from real UI screenshots that resemble the synthesized examples. During the style merge, the GANSpiration-based approaches alter the layout and/or the details of the original input, leveraging StyleGAN's architecture. Note that although user interface design artifacts can be represented in many formats (e.g., a tree of UI components~\cite{Wu2021}), we considered the UI screenshot images in our approach since they are one of the most frequent types of inspirational sources used by the designers. When investigating and evaluating GANSpiration, we pose the following research questions:

\begin{enumerate}[noitemsep, label={\textbf{RQ{{\arabic*}}}:}]
    \item How do GANSpiration-based approaches compare with random examples and similarity-based examples in quantitative metrics indicating the ability of inspiration support?
    \item How do UI/UX practitioners perceive the output of GAN-Spiration-based approaches in comparison to random examples and similarity-based examples?
\end{enumerate}

To answer these questions, we first developed our approach which extends the StyleGAN architecture trained on a large-scale dataset including 58,040 screenshots of Android applications. We then proposed two quantitative measurements for evaluating the ability of inspiration support of a set of UI images: (1) \textit{similarity} of the images to the input image and (2) \textit{diversity} of the set of UI images. We found that the GANSpiration-based methods provide much more diverse design examples than a similarity-based method, and at the same time they provide more similar examples to the input image than a random example selection approach, indicating a balance between diversity and relevance. Through a user study with five professional UI/UX practitioners, we found that the participants perceived the GANSpiration-based methods as a viable way to gain inspiration to modify a UI design. Overall, our work contributes a novel and promising approach in which a style-based generative machine learning technique is applied in the context of inspiration and creativity support in user interface design. We believe that the ideas presented in our approach will encourage and influence more research efforts towards the pragmatic use of generative machine learning models in the creative, yet constraint, design tasks.
\section{Related Work and Background}
Our work is most closely related to previous studies that focused on (1) design inspiration, (2) techniques for managing design artifacts, and (3) generative machine learning models and StyleGAN in particular. We briefly review each group of literature in the following sections.

\subsection{Design Inspiration}
Thrash et al.~\cite{Thrash2003,Thrash2014} were among the first who empirically studied inspiration as a psychological construct. They have identified that human inspiration is categorized by motivation (i.e. goal-oriented self-initiation), evocation (i.e. an impulsive reaction to stimuli), and transcendence (i.e. feeling of gaining superior ideas that are ``more elegant or novel than those generated willfully''). 

The problem of inspiration has been then investigated in a wide design community, beyond user interaction design. These previous studies were mostly conducted from the perspectives of how designers get access to and use existing design artifacts. For example, focusing on knitwear design, Eckert and Stacey~\cite{Eckert2000} have identified that designers used a wide variety of sources of inspiration, including artifacts with intriguing shapes, patterns, and colours, as well as their own previous design, to not only concretize the otherwise abstract design ideas, but also to create ``shortcuts'' to help them recall and communicate using these visuospatial ``chunks;'' i.e. inspirational sources served as ``a language of design.''

In the HCI community, researchers have explored ways how industrial and user interaction designers get inspired by existing design artifacts. For example, Bonnardel~\cite{Bonnardel1999} has identified that, in the context of product design, ``the emergence of new ideas results from analogy-making.'' From an in-depth interview study with web, graphic, and product designers, Herring et al.~\cite{Herring2009CHI} identified the common approaches they used and the challenges they faced when retrieve, store, and disseminate design examples. Based on a glossary of design ideation methods, Gon\c{c}alves et al.~\cite{Goncalves2014} have also conducted a survey with students and professional industrial designers to understand their sources and methods of inspiration. They found that, comparing to students, professional industrial designers adopted a wider variety of inspirational approaches.

The literature has also identified several problems and issues about the common inspirational methods. Notably, many studies have pointed to the fact that over-exposure to a homogeneous set of design examples may result in ``design fixation,'' which will limit the inspirational power of the examples and result in less creative ideas~\cite{Jansson1991, Marsh1996, Herring2009CHI}. Particularly, Marsh et al.~\cite{Marsh1996} identified that exposing to a greater number of examples that share common critical characteristics would increase the fixation issue. The timing of the example exposure can affect the quantity and quality of ideas as well. For example, Siangliulue et al.~\cite{Siangliulue2015} found that receiving examples when their participants seemed to have run out of ideas have allowed the participants to produce a larger number of ideas, whereas explicitly requesting examples when needed have allowed the participants to produce more novel ideas. In this study, we build on this body of literature to investigate techniques for supporting effective inspiration in user interface design, while avoiding design fixation.

\subsection{Managing UI Design Artifacts}
While abundant recent work focused on extracting UI elements, including their hierarchical design information, from design artifacts such as mockups (e.g. \cite{Chen2018ICSE,Beltramelli2018,Swearngin2018,Moran2018,10.1145/3368089.3417940,10.1145/3180155.3180240}), they are not directly related to the objective of providing inspirational design examples. So we omit the detailed review of this body of literature here. In this section, we focus on reviewing related work that investigates the management of UI design artifacts for the purpose of design inspiration.

Towards this direction, some previous studies have focused on techniques that retrieve UI design examples based on an input UI screenshot. For example, Lee et al.~\cite{Lee2010} have proposed an ``Adaptive Ideas'' web design tool, which allows users to view examples similar to their current design work. In their tool, the users could control the dimensions (including background color, primary font, number of columns, and visual density) used to compare design similarities. Similarly, Behrang et al.~\cite{Behrang2018} proposed a technique that combined keyword search and image-based search to retrieve apps (along with their code) with similar screenshots as an input design. Hashimoto et al.~\cite{Hashimoto2005} have also introduced a technique that aims to help inexperienced designers retrieve similar design examples based on an input in the form of sketch or wireframe. Ritchie et al.~\cite{Ritchie2011} have proposed a design exploration tool that allows its users to query design examples by descriptive text including color keywords or style terms; the tool can also search by style similarity.

The recent development of deep neural networks has enabled more powerful techniques for similarity-based design example retrieval. For example, Huang et al.~\cite{Huang2019} introduced Swire, a sketch-based neural network-driven technique for retrieving user interface designs. The core component of Swire is a deep convolutional neural network (CNN)~\cite{lecun1995convolutional} that calculates the ``embedding'' (i.e. a numerical representation) of a design artifact (e.g. a sketch or a screenshot). Once trained, Swire could retrieve UI design artifacts that are similar to an input sketch or screenshot. More recently, Bunian et al.~\cite{10.1145/3411764.3445762} proposed VINS to retrieve the most structurally similar UI screenshots to the input using object detection models to identify the UI components of the UI screenshots or wireframes. Based on the components and their layout, an image retrieval model helps to find similar UI screenshots in the reference dataset.

Notably, most previous studies relied on similarity when retrieving design artifacts, which can result in design fixation and may hinder the creative design process. Our study addresses this issue by focusing on a style-based generative approach that balances the targeted and serendipitous aspects of design artifact retrieval for inspiration.

\subsection{StyleGAN and its Application on UI Design}
StyleGAN, or Style-Based Generative Adversarial Network~\cite{karras2018stylebased}, extends the traditional GAN~\cite{Goodfellow2014GAN} architecture on a style-based generator model inspired by the style transfer literature. GANs typically include two machine learning models that are trained at the same time: a \textit{generator} trained to synthesize data points (e.g., images) that resemble those in the original dataset and a \textit{discriminator} that learns to classify if an input image is synthesized by the generator. Once trained, the generator can be used to synthesize images from an input vector in the latent space. Based on the GAN architecture, StyleGAN proposes a style-based generator that focuses on explicitly transferring `styles' on an image at different resolution levels during the synthesis process. This results in a synthesized image with one input vector, or ``latent code'', dominating its overall features and the other latent code contributing mostly to the details of the image~\cite{karras2018stylebased}. 

While GAN and StyleGAN have obtained great attention given their capacity for generating high-resolution and realistic-looking images, their application for UI design is still in its infancy. The only previous work that used generative models for providing UI examples is a very recent study done by Zhao et. al.~\cite{zhao2021guigan}. They developed a technique to generate UI structures and reused UI components collected from existing mobile apps to fill in the generated structure in order to create UI examples. Their study only focused on the quality of the generated UIs, in terms of metrics such as color harmony and structure rationality. Additionally, their evaluation was not done with professional designers who are familiar with real-life design practices. Instead, we aim to understand the ability of the generative models in providing inspirational design support. Our techniques also address style-based design transformation, which is not explored in the literature. 

StyleGAN has been improved since its publication. While our work is based on the original StyleGAN work, we expect the performance of our approach can be enhanced by using more advanced generative models, such as StyleGAN2~\cite{Karras2019stylegan2}. Our contribution, however, is not on using the most recent models, but instead on illustrating the potential of applying this line of work for a novel but important problem, i.e., generating design examples with high diversity and relevance for effective inspiration.
\section{GANSpiration System Architecture}
\label{sec:architecture}

\begin{figure*}[t]
    \centering
        \includegraphics[width=\textwidth]{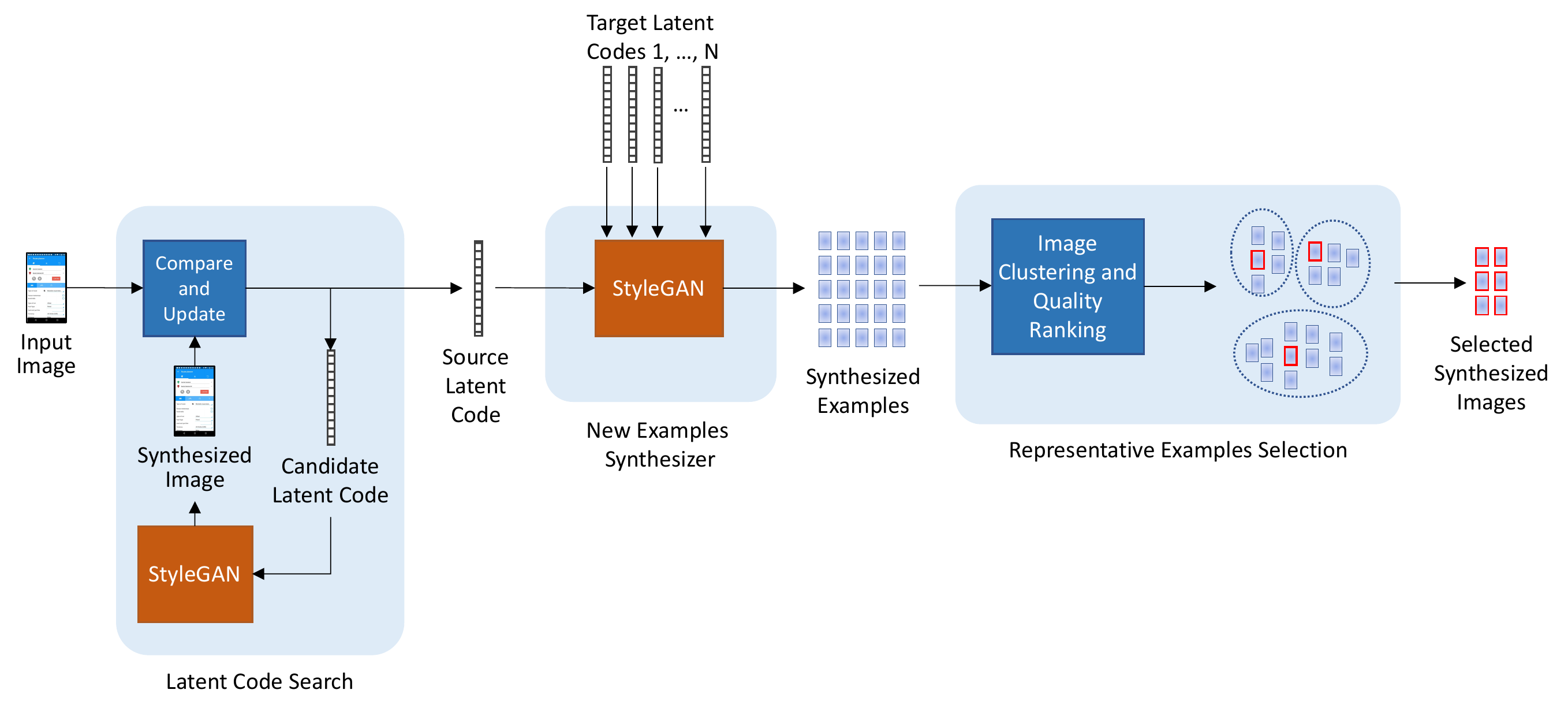}
        \caption{Overview of GANSpiration. GANSpiration takes a preliminary design image as input then transforms it into a latent code. The style of the original input is then merged with a set of target latent codes to synthesize new examples. Image encoding and clustering methods are then used to find the representative examples as output. While not illustrated in this diagram, the output of the synthesized examples can be additionally used to search real UI screenshots for inspiration.}
        \Description{A diagram illustrating the architecture overview of GANSpiration.}
        \label{fig:overview}
\end{figure*}

The interaction between the designer and GANSpiration is initiated when the designer has a preliminary design artifact (e.g., a UI mockup image) at hand, related to their design task. The designer sends this image as an input to GANSpiration, which will first be encoded into a latent code (i.e., a high-dimension vector in the latent space). This latent code is then merged with other latent codes, either randomly generated or obtained from other UI images, to synthesize a unique set of new example images. From this set, the system selects the most representative example images and displays them to the designer. The designer can also configure GANSpiration to return the real UI screenshots from the database that are the closest to the generated results. As such, the GANSpiration architecture is primarily comprised of three components to achieve its key functionality: (1) a latent code search component, (2) a new examples synthesizer, and (3) a representative examples selection component. The overall architecture is illustrated in Figure~\ref{fig:overview}. Below, we first describe each component in detail. Then we describe the process used to train the StyleGAN that supports these components.

\subsection{Component 1: Latent Code Search}
\label{sec:comp1}
StyleGAN relies on a condensed representation of images called latent codes, i.e., high dimensional vectors (512 dimensions in our case), to perform image synthesis. The input images of GANSpiration, therefore, need to be first encoded from their original format to the latent space that corresponds to a trained StyleGAN model. This is done through Latent Code Search, which is built upon the work of StyleGAN-Encoder~\cite{latent_code_search}. This component searches the optimal latent code of an input image through a gradient descent update based on the difference between the synthesized image and the input image evaluated using a perceptual model~\cite{Zhang_2018}. The latent code obtained is then returned to represent the input image. 

\subsection{Component 2: New Examples Synthesizer}
\label{sec:comp2}
New Examples Synthesizer merges the style of the input image as source with a set of target images or latent codes to produce a new set of examples. In this process, the source image latent code is obtained from the previous step, while the target latent codes can be any vectors from the latent space depending on the configuration of GANSpiration. For example, it can use the latent codes obtained from a set of existing target images or latent codes that are directly sampled from the latent space. 

We set StyleGAN to include eight levels of image resolutions during the synthesis process~\cite{karras2018stylebased}; each resolution level is comprised of two style input locations into which we feed the source and target latent codes. GANSpiration merges images by replacing the latent code of the source image for certain style input location(s) with the target latent code. We iterate all the consecutive style input locations, resulting in $136$ images generated for each pair of source and target inputs. This set of output covers all possible granularity levels of influence of the target image (i.e., from the structural level to the detail level)  on the source image.

\subsection{Component 3: Representative Examples Selection}
\label{sec:comp3}
In the previous step, we synthesized a large number of new examples for each source-target pair, which can be overwhelming for the designers to examine. Moreover, since our style merging process is very fine-grained, it also introduces redundancies in the set of synthesized examples. In this component, we apply a clustering method to pick a smaller sample of representative images from the set of generated images. In particular, we adopt the DBSCAN method~\cite{birant2007st} (with threshold $\epsilon$ as 0.9) to cluster the images through perceptual similarity~\cite{Zhang_2018} calculated between any two images from the synthesized example set. Within each cluster, we selected the image that the discriminator of the trained StyleGAN considered as the closest to the real UI screenshot as the representative example. The output of this component is a collection of synthesized images that are sufficiently different from each other and contain either coarse or detailed design aspects of the input image.

\subsection{Training StyleGAN for GANSpiration}
To use the capacity of StyleGAN in the context of UI design, we need to retrain the StyleGAN model with a dataset of user interface design artifacts. In this section, we describe the dataset we used and our training process of StyleGAN for GANSpiration. The trained StyleGAN model was then used in the components described above.

\begin{figure*}[t]
    \centering
    \includegraphics[width=0.65\textwidth]{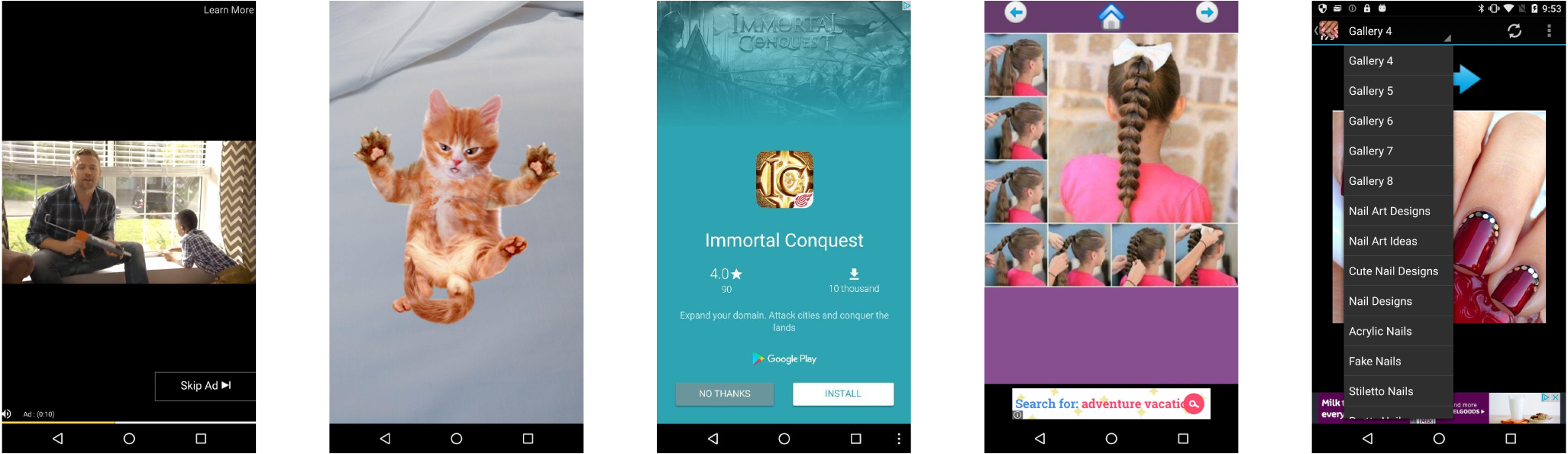}
    \caption{Samples of UI screenshots removed because they only have one or two unique component types. Most of the screenshots in this category contain splash screen, video screenshot, full-screen image, and web view in their design, adding unnecessary visual complexity.}
    \Description{Five images showing UI screenshots that each contained: (1) a video screen capture, (2) an image of a cat, (3) an app installation screen, (4) several images showing steps for making a braided hairstyle, and (5) an image showing a fingernail style.}
    \label{fig:removed_examples}
\end{figure*}

\begin{figure}[t]
    \centering
    \includegraphics[width=0.9\columnwidth]{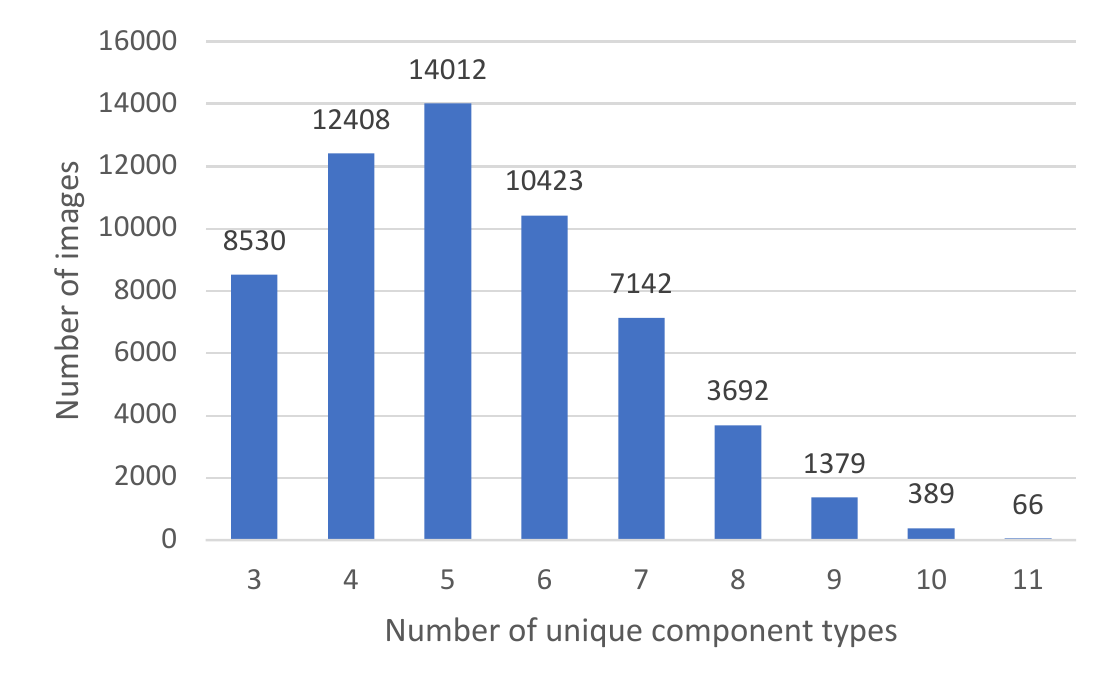}
    \caption{The distribution of the number of unique component labels in the UI screenshots in our dataset.}
    \Description{A bar-chart showing the number of images for each category of number of unique component types. Detailed data: 8530 images contained three unique component types, 12408 image contained four unique component types, 14012 image contained five unique component types, 10423 images contained six unique component types, 7142 images contained seven unique component types, 3692 images contained eight unique component types, 1379 images contained nine unique component types, 389 images contained ten unique component types, and 66 images contained 11 unique component types.}
    \label{fig:unique_labels_distribution}
\end{figure}

\subsubsection{Dataset}
We used the Rico dataset~\cite{Deka2017Rico} for StyleGAN training. It contains 66,261 unique UI screenshots of Android apps and serves as one of the largest repositories of mobile app designs to date. The dataset includes a diverse set of UI screenshots with varied complexity that contains various types of UI components. In order to obtain a high-quality dataset for inspiration purposes, we removed UI screenshots that have only one or two unique component labels from the Rico dataset. This is because we found in a manual inspection that the UIs with less than three unique component types usually do not contain enough interaction elements to support inspiration or benefit from our approach. They also often introduce unnecessary visual complexity that only exists in singular instances (e.g., a splash screen, video screenshot, advertisement, or full-screen image, see Figure~\ref{fig:removed_examples}) that affects the training performance; in other words, these instances make it difficult for StyleGAN to generate similar components and even if generated, the visual presentation of those components are too uncommon to be useful for inspiration. We used the UI view hierarchy data in the Rico dataset to calculate the number of unique component types in each screenshot. In total, we removed 8,221 images that had only one or two unique component types, resulting in 58,040 images in the dataset. Figure~\ref{fig:unique_labels_distribution} shows the distribution of the number of unique component labels in the UI screenshots in the dataset. We then resized each image into $1024\times1024$ for the training purpose.

\subsubsection{Training process}
We built upon the official TensorFlow implementation of StyleGAN\footnote{https://github.com/NVlabs/stylegan} and used our preprocessed dataset to train the model. While we did not perform a formal hyperparameter search, we explored a few changes of hyperparameters reported in Karras et al.~\cite{karras2018stylebased}, including the initial learning rates of the generator and the discriminator, the number of times the discriminator was trained per generator iteration, and the number of minibatches to run before adjusting the training parameters. We eventually used the following hyperparameters because they achieved the best performance according to the Fr\'echet inception distance (FID)~\cite{Heusel2017FID} in our exploration: for both the generator and the discriminator, the learning rate was set to 0.0015 for resolution levels equal or less than $128\times128$, 0.002 for resolution levels $256\times256$ and $512\times512$, and 0.003 for $1024\times1024$; the discriminator was trained at the same frequency as the generator; and training parameters were adjusted after every four minibatches. Additionally, mirror data augmentation used in the original StyleGAN work was not enabled during training due to the unsymmetrical nature of UI images. We used a server that contained four NVidia V100SXM2 GPUs to train the model. Training terminates when the FID value increases (i.e., the generation quality deteriorates) for three consecutive iterations. The entire training process lasted 162 hours. The best performance measured with FID was achieved at 42.91. Although this performance is not ideal compared to the typical face generation tasks, we found that the generated images can already represent certain layout features and visual details that can help design inspiration. This difference in performance may due to the extraordinary complexity and diversity of screenshots of UI design. It is worth noting that our focus and contribution in this paper are not to achieve a higher performance in the generative model. Instead, we focus on evaluating the potential of this line of techniques for supporting the challenging task of design inspiration, even with less-than-perfect image generation.

\section{Sample Usage Scenario}
The main purpose of GANSpiration-based approaches is to inspire designers with a set of UI design examples that are diverse enough to help avoid design fixation, while relevant to the designer's work at hand. To illustrate the design process with and without the support of GANSpiration, we describe a sample usage scenario from a UI designer's perspective. This usage scenario is created based on our informal discussion with several designers and is used to guide the design of our evaluation studies.

Daphne worked as a UI/UX designer for six years at a large company. She was recently positioned to lead the design of a revamped version of the company's main product. Before Daphne adopted the GANSpiration-based tool, she used to rely on two main ways to draw inspiration and reference from other systems and design examples during the ideation phase (sometimes with the users and/or the development team). First, Daphne frequently visited her favorite design gallery platforms, Dribbble~\cite{dribbble} and Behance~\cite{behance}, either searching for examples for a certain type of functions and components or simply fumbling through the galleries until ideas hit her. These galleries contain collections of screenshots of design examples in various application domains, with drastically different design styles. So, although she felt that they are useful to help her think out of the box, she sometimes found those examples distracting and the whole process ineffective. Second, Daphne also had access to a similarity-based design example retrieval tool that her company procured. This tool would provide her with a collection of UI design screenshots that looks similar to a preliminary mockup that she created. She found this tool particularly useful in helping her to be focused on the design elements presented in her mockups. However, the design examples suggested by the tool often look very similar. As a result, Daphne often found herself stuck in a certain way that a UI is `supposed' to look like and lost her creative power.

Daphne found the GANSpiration-based tool achieving a good balance between the two inspirational approaches she used before. At the beginning of the design revamping work, Daphne fed the screenshots of the old design to the tool and received a collection of synthesized design examples that resemble either the structure or the details of the original design but each had unique aspects that Daphne found interesting. She was able to create an initial revamped design mockup by combining several elements and aspects from the synthesized images. Daphne found that GANSpiration can effectively support communicating design ideas too. During a co-design session with two power users of their product, Daphne received a suggestion of changing the presentation of a list of items on the mockup. Examining the GANSpiration output of the mockup, Daphne got the idea of either adding an icon to each list item (i.e., a detailed change) or changing the list into a wall of cards (i.d., a structural change). Daphne picked a few examples synthesized by the tool to show to the users to illustrate her ideas and eventually settled on the wall-of-cards concept. Overall, Daphne was glad that she could easily obtain a collection of design examples in various occasions that were not only diverse enough to allow her to explore different design ideas but also relevant to her work at hand.

\section{Quantitative Evaluation}
We conducted an experiment that focused on evaluating the generative methods against two other techniques that suggest design examples (i.e., random suggestion and similarity-based suggestion). The evaluation is based on the usage scenario we described above. The random suggestion baseline resembles the serendipity-based inspirational process that relies on exploration and encountering, while the similarity baseline reflects the similarity-based design example retrieval approach. While we used two quantitative metrics (i.e., similarity and diversity) for measuring the ability of inspiration support of the system outputs, we also focus on discussing our observations and insights on how the outputs might have contributed to the metrics. 

\subsection{Data sampling}
\begin{figure*}[t]
    \centering
    \includegraphics[width=0.87\textwidth]{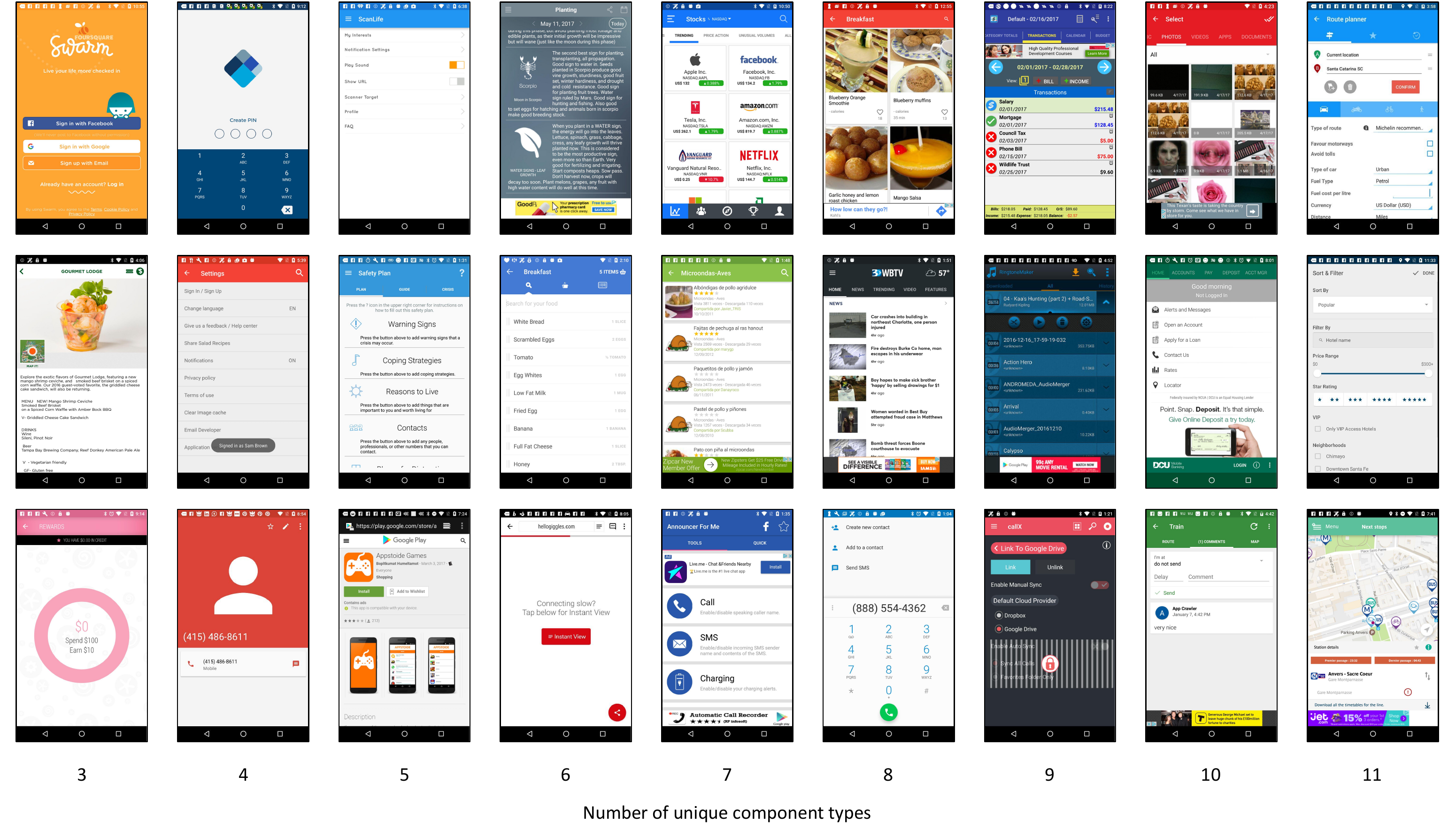}
    \caption{Sample of UI screenshots used in the evaluation study, ordered by the number of unique component types.}
    \Description{The figure shows 27 UI screenshots organized in a grid of three rows by nine columns. Each column indicate a category of a number of unique component types, from 3 to 11.}
    \label{fig:all-sample}
\end{figure*}

To sample a diverse set of UI screenshots as inputs to GANSpiration, we divided the Rico dataset according to the number of unique component types on the UIs. We considered the number of unique component types as an indicator of the complexity of the UI, thus the complexity of the design task. Our dataset contained UI screenshots that contain 3 to 11 unique component types, resulting in nine unique groups (see Figure~\ref{fig:unique_labels_distribution}). We randomly selected three UI screenshots from each group, resulting in 27 screenshots as inputs in the evaluation scenario; these images are shown in Figure~\ref{fig:all-sample}. This strategy ensures the coverage of different levels of complexity in the UI inputs, thus the coverage of the complexity of the design task.

\subsection{Quantitative metrics}
\label{sec:metrics}
We derived two metrics to measure the ability of inspiration support of a set of UI images. These metrics are inspired by previous work about inspiration in design~\cite{Herring2009, Lee2010} and collectively focused on both targeted and serendipitous inspiration. Both metrics rely on a measure of distance between images, particularly a measure based on a perceptual distance model aimed to approximate human visual perception; this measure is the same as the one used in the clustering task in the representative examples selection component of GANSpiration (see section~\ref{sec:comp3}). 

In the following equations, $dist(A,B)$ denotes the perceptual distance between images $A$ and $B$, ranging from 0 to 1 inclusively, calculated using Zhang et al.'s technique~\cite{Zhang_2018}; $E_i^{(D)}$ ($i=1,2,...,n$) denotes the $i$th output example image for an input design $D$; and $O^{(D)}=\{E_1^{(D)},E_2^{(D)},...E_n^{(D)}\}$ denotes the set of output images for an input $D$. The two metrics we used are:

\begin{itemize}
    \item \textbf{Similarity} of the suggested examples to the input design. This metric is double-sided. A sufficient similarity may indicate the relevance of the suggested examples, which is important to provide targeted inspiration. However, a high similarity indicates the potential of design fixation. We use the mean similarity (i.e., the complement of the perceptual distance) between the output images and the input image to evaluate the overall similarity of the output examples to an input design:
    \begin{equation*}
        Sim(O^{(D)})=1-\frac{1}{n}\sum_{i=1}^n dist(D, E_i^{(D)})
    \end{equation*}
    \item \textbf{Diversity} of the suggested examples indicates how varied the outputs are, given an input design. It plays an important role in preventing fixation. We use the mean of pairwise distances among all output examples of an input design to evaluate the diversity of the output set:
    \begin{equation*}
        Div(O^{(D)})=\frac{1}{n(n-1)/2}\sum_{j=i+1}^n\sum_{i=1}^{n-1} dist(E_i^{(D)}, E_j^{(D)})
    \end{equation*}
\end{itemize}

\subsection{Experimental design}
\begin{figure*}[t]
    \centering
    \includegraphics[width=0.82\textwidth]{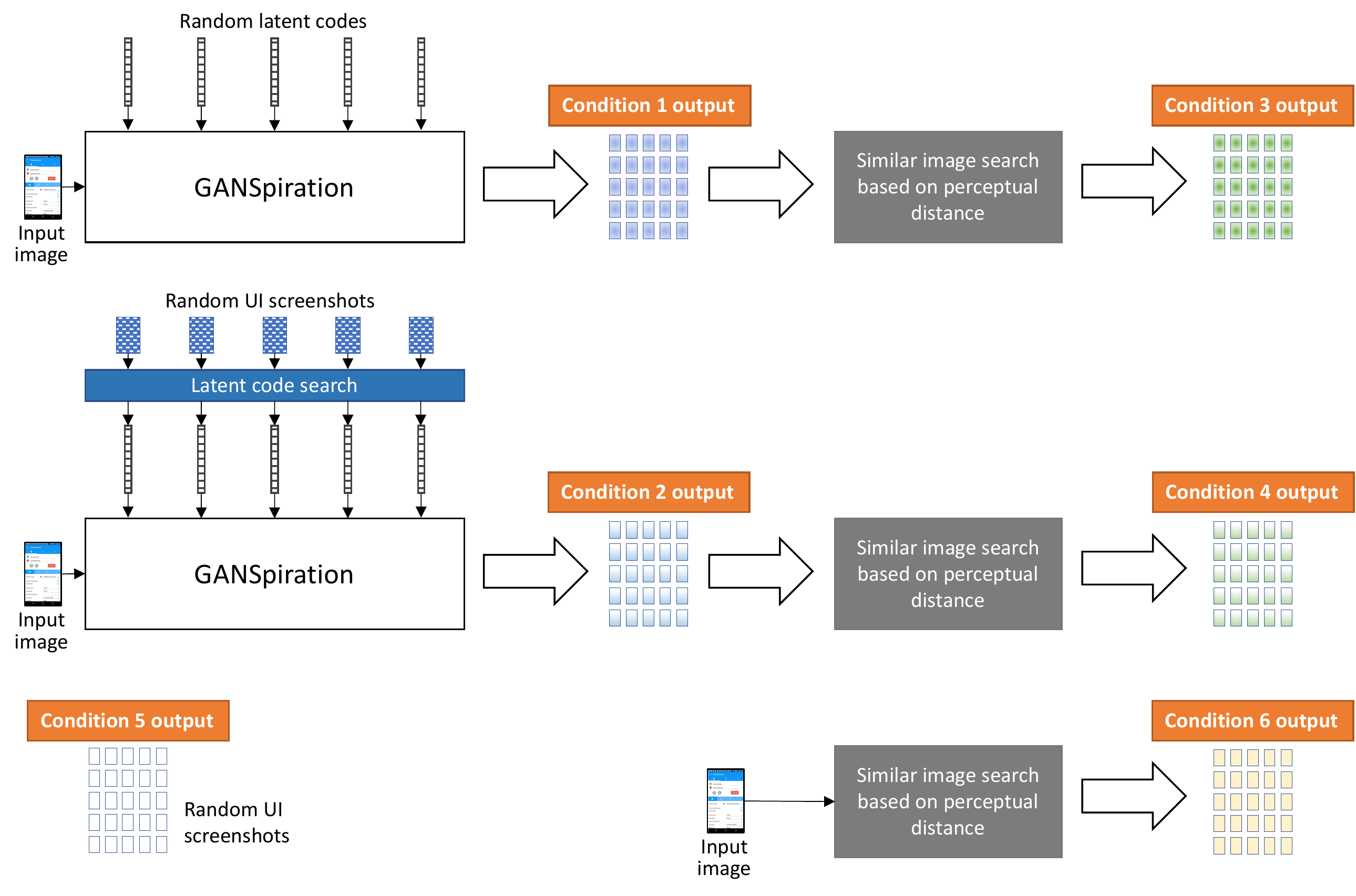}
    \caption{The six experimental conditions. The condition outputs were then used to calculate the metrics and in the user studies.}
    \Description{A diagram illustrates the six experimental conditions described in section 5.3.}
    \label{fig:conditions}
\end{figure*}

In the experiment, we considered the following conditions for suggesting a set of images for inspiration. Particularly, conditions 1, 2, 3, and 4 are four variants of the GANSpiration method. Figure~\ref{fig:conditions} summarizes these conditions.

\begin{figure}[t]
\centering
\begin{subfigure}[b]{0.4\textwidth}
    \centering
    \includegraphics[width=\textwidth]{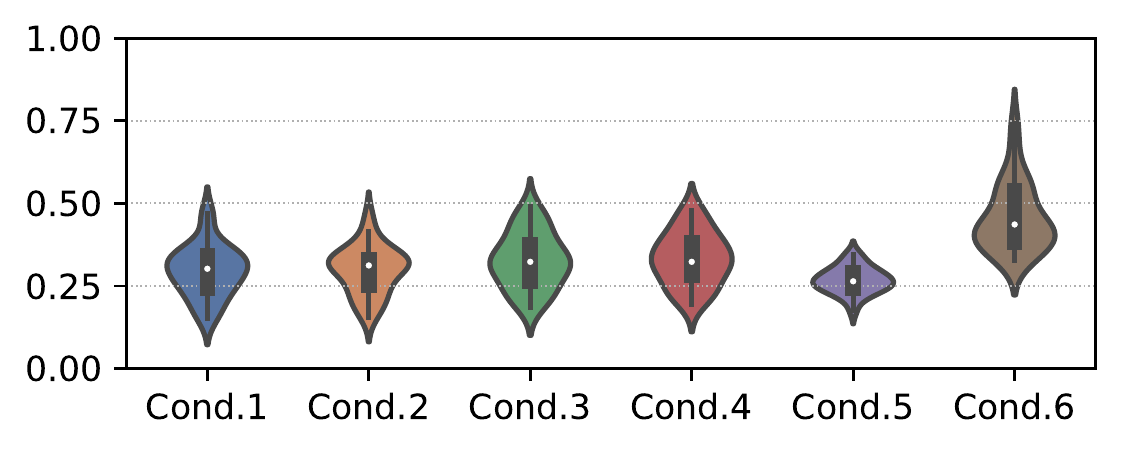}
    \caption{similarity}
    \label{fig:distribution-metrics-similarity}
\end{subfigure}
\hspace{30pt}
\begin{subfigure}[b]{0.4\textwidth}
    \centering
    \includegraphics[width=\textwidth]{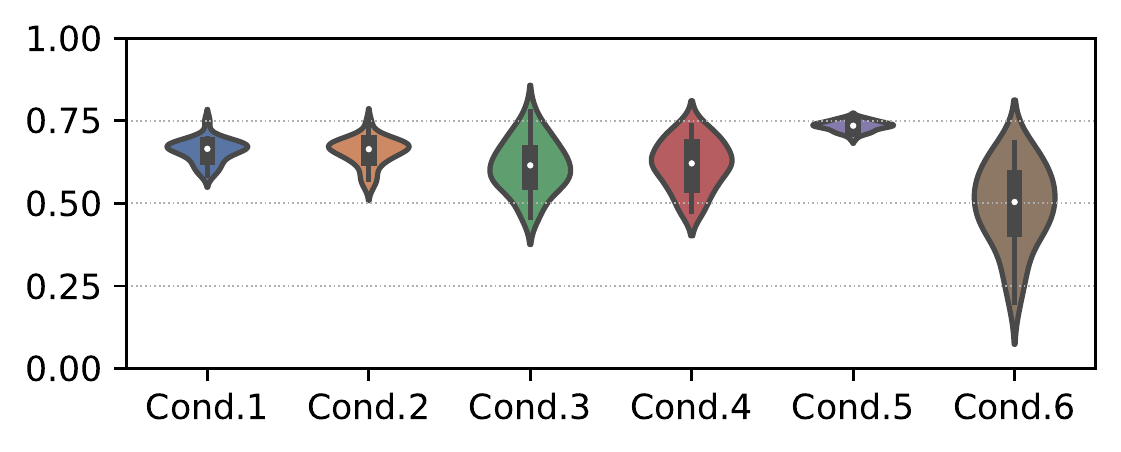}
    \caption{diversity}
    \label{fig:distribution-metrics-diversity}
\end{subfigure}
\caption{Distribution of the similarity and diversity metrics on the experimental conditions.}
\Description{Two violin charts showing similarity and diversity distributions for each experimental condition. For similarity, Condition 6 has a higher median (around 0.45) than the other conditions (around 0.3). For diversity, Condition 5 has a higher median (around 0.75) and a narrower distribution, Condition 6 has a lower median (around 0.5) and a wider distribution, and all other conditions have a median of around 0.65.}
\label{fig:distribution-metrics} 
\end{figure}

\textbf{Condition 1}. 
In this condition, we used the trained StyleGAN model to merge the input image with five random latent codes, each generated a set of examples according to the GANSpiration architecture (see Section~\ref{sec:architecture}). We then directly used these examples as the output. %1-1
    
\textbf{Condition 2}.     
This condition is similar to Condition 1; however, instead of using five random latent codes, we randomly selected five images from the prepossessed dataset and obtained their latent codes for style merging. The examples generated were then used as the output. Conditions 1 and 2 are created to evaluate two different variants for using directly generated images for inspiration. %1-2

\textbf{Condition 3}.
In this condition, we first obtained the output of Condition 1. Then for each generated output image, we searched for the most similar image, using the perceptual distance described in section~\ref{sec:metrics}, from a dataset of real UI screenshots. Because of the computational cost of this search, we used a smaller dataset created by Huang et al.~\cite{Huang2019} as the search space; it contained 2201 high-quality UI screenshots sampled from the Rico dataset. Real UI screenshots obtained from this search were then used as the output. %2.1-1
    
\textbf{Condition 4}.
This condition is similar to Condition 3 except that the initial generated images for search were obtained from Condition 2 instead of Condition 1. Conditions 3 and 4 are created to examine the effects of how realistic the design examples are on inspiration. %2.1-2

\textbf{Condition 5}.
In this condition, we randomly selected 25 images from the prepossessed dataset as the output.
    
\textbf{Condition 6}.
In this condition, we performed a search on a subset of Rico dataset created by Huang et al.~\cite{Huang2019} to obtain the most similar images to the input image, based on the perceptual distance described in section~\ref{sec:metrics}. The top 25 similar images were used as the output.

\subsection{Results}

\begin{figure*}[t]
    \centering
    \includegraphics[width=0.9\textwidth]{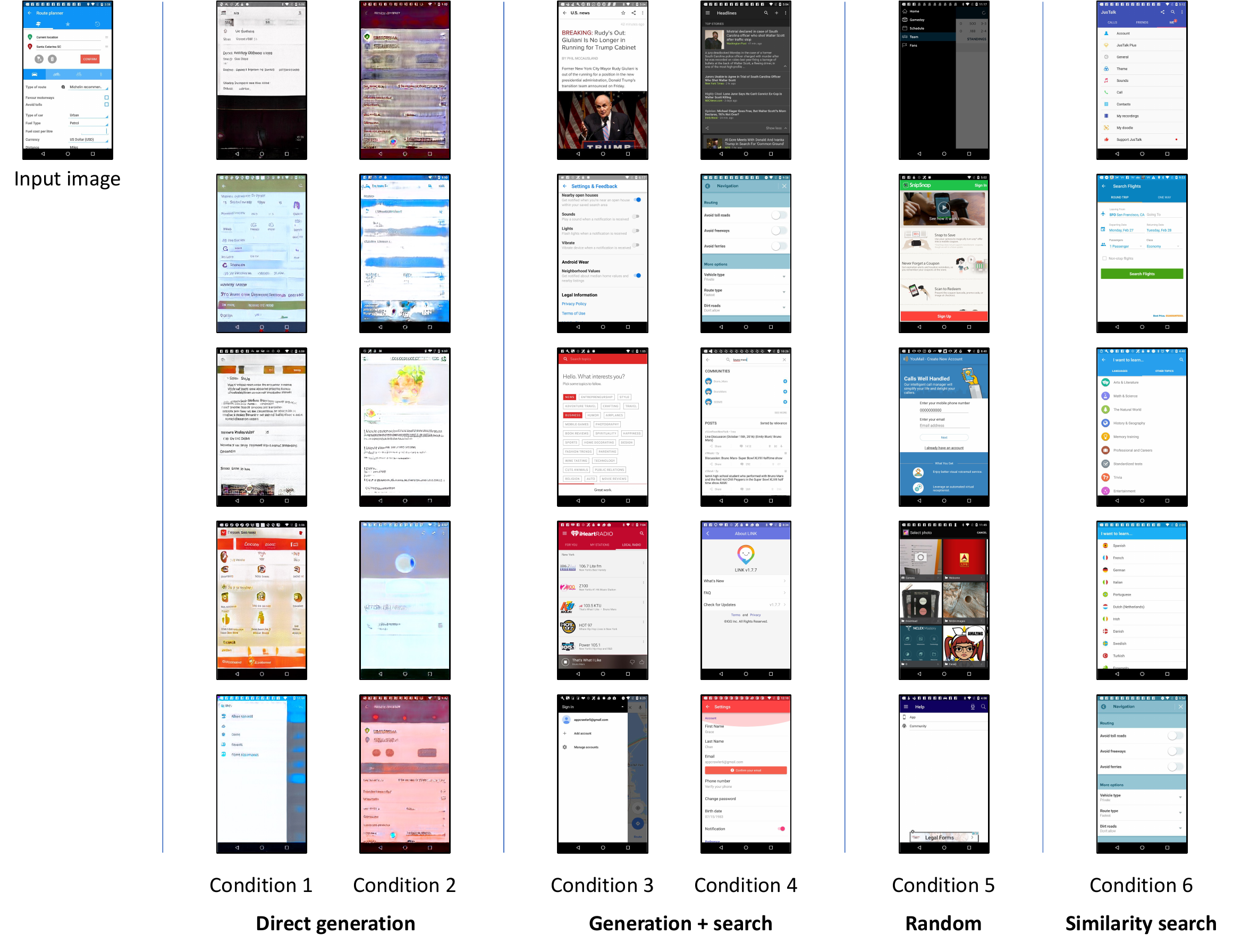}
    \caption{Samples of five outputs from each experimental condition for the input image. Samples from Condition 3 are matched with those from Condition 1, and samples from Condition 4 are matched with those from Condition 2.}
    \Description{An input UI screenshot is shown on the left most side of the figure. Then next to the input screenshot, there are 30 UI images organized in a grid of five rows by six columns; each column indicates an experimental condition, from Condition 1 to Condition 6.}
    \label{fig:condition-output-comparison}
\end{figure*}

Among all the sampled inputs, both Condition 1 and Condition 2 have resulted in an average of 19 output images ($SD=7.2$ and $5.15$, respectively). A comparison of samples of five outputs from each experimental condition for one input image is shown in Figure~\ref{fig:condition-output-comparison}. Figure~\ref{fig:distribution-metrics} shows the distributions of the similarity and diversity metrics on the six conditions. Kruskal-Wallis tests indicated statistically significant differences among the experimental conditions with respect to both metrics ($p<0.001$). We then conducted posthoc pairwise analyses using the Mann-Whitney U test with Bonferroni correction to identify the conditions that contributed to the difference; Bonferroni correction was used to address multiple comparisons. The posthoc analyses revealed the following results.

\begin{itemize}
    \item The GANSpiration-based methods (Conditions 1, 2, 3, and 4) have resulted in significantly lower similarity ($p<0.001$) than Condition 6 (i.e., search-based approach). Conditions 3 and 4 (i.e., generation + search) also resulted in significantly higher similarity ($p<0.05$), thus relevance, than random examples (Condition 5).
    \item While the GANSpiration-based methods (Conditions 1, 2, 3, and 4) have resulted in significantly lower diversity ($p<0.001$) than Condition 5 (i.e., random examples), they have also achieved significantly higher diversity ($p<0.01$) than similarity-based approach (Condition 6).
\end{itemize}

\begin{figure*}[t]
\centering
\begin{subfigure}[b]{0.9\textwidth}
    \centering
    \includegraphics[width=\textwidth]{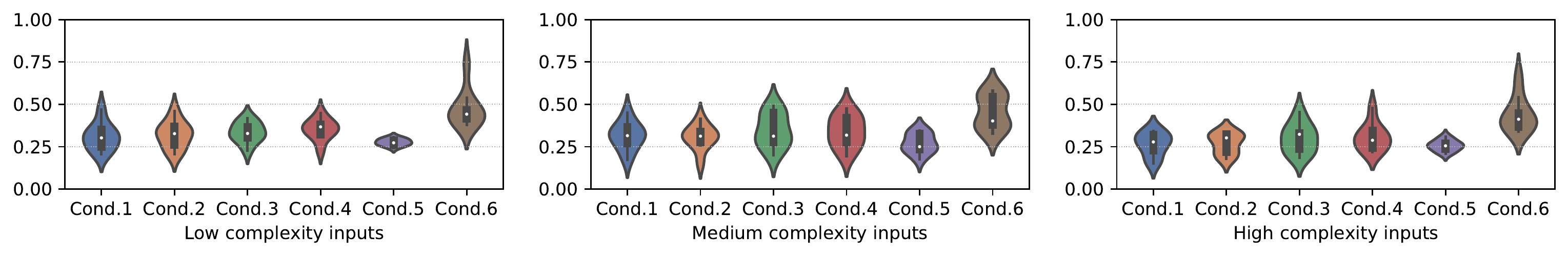}
    \caption{similarity by input complexity}
    \label{fig:distribution-metrics-similarity-by-complexity}
\end{subfigure}
\begin{subfigure}[b]{0.9\textwidth}
    \centering
    \includegraphics[width=\textwidth]{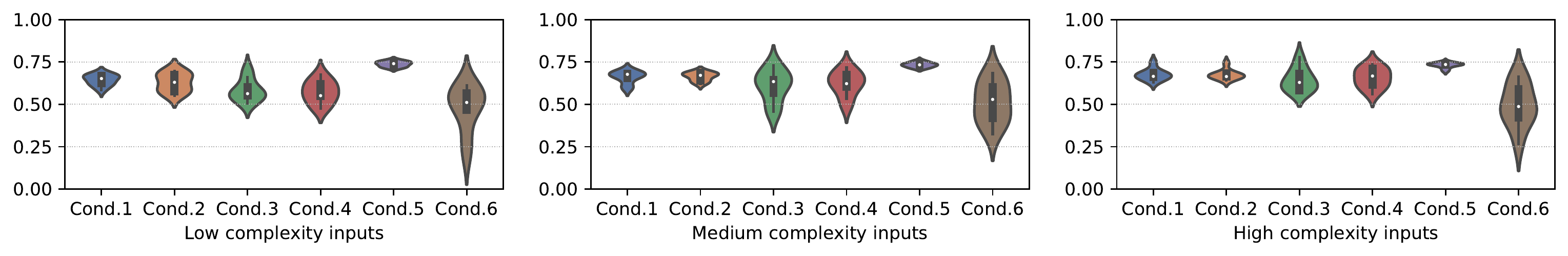}
    \caption{diversity by input complexity}
    \label{fig:distribution-metrics-diversity-by-complexity}
\end{subfigure}
\caption{Distribution of the similarity and diversity metrics on the experimental conditions, analyzed by input complexity.}
\Description{Two sets of three violin charts showing similarity and diversity distributions for low, medium, and high complexity inputs, for each experimental condition.}
\label{fig:distribution-metrics-by-complexity} 
\end{figure*}

To understand how the complexity of the input UI design can influence the similarity and diversity metrics of the six experiment conditions, we separated the 27 sampled inputs into three groups: (1) \textit{Low complexity inputs} contain less than six unique component types. In our sample, they often represent login screens, setting menu screens, or screens that communicate a single piece of information (see Figure~\ref{fig:all-sample}). (2) \textit{Medium complexity inputs} contain between six to eight unique component types. In our sample, they often represent screens that contain heterogeneous information (usually presented in lists or cards) or screens that provides multiple options to users. (3) \textit{High complexity inputs} contain more than eight unique component types. In our sample, they often represent screens that contain complex interaction mechanisms, including tabs, multiple options on list items, maps or complex forms. Figure~\ref{fig:distribution-metrics-by-complexity} presents the distributions of the similarity and diversity metrics in each input complexity group. Kruskal-Wallis tests indicated statistically significant differences among the experimental conditions with respect to both metrics in all three groups ($p<0.05$). Table~\ref{tab:mean-difference-similarity-by-complexity} and Table~\ref{tab:mean-difference-diversity-by-complexity} present the posthoc analysis results based on pairwise Mann-Whitney U tests with Bonferroni correction. Results indicated that certain GANSpiration-based approaches (particularly Conditions 1 and 2) achieved a significantly lower similarity than Condition 6 (i.e., search-based approach) for low and high complexity inputs, but not in medium complexity inputs. Additionally, for high complexity inputs, the GANSpiration-based approaches can achieve a similar level of diversity to random suggestions (Condition 5), and significantly higher diversity than the search-based suggestions (Condition 6). 

\begin{table*}[t]
\centering
\small
\caption{Differences of medians (row minus column) on the similarity metric among the six conditions, by input complexity.}
\label{tab:mean-difference-similarity-by-complexity}
\begin{subtable}[c]{0.3\textwidth}
    \centering
    \subcaption{Low complexity inputs}
    \resizebox{\textwidth}{!}{
    \begin{tabular}{crrrrr}
    \toprule
    & \multicolumn{1}{c}{C.2} & \multicolumn{1}{c}{C.3} & \multicolumn{1}{c}{C.4} & \multicolumn{1}{c}{C.5} & \multicolumn{1}{c}{C.6} \\
    \midrule
    Cond.1 & -0.03 & -0.03 & -0.07 & 0.03 & *-0.14 \\
    Cond.2 &       & -0.00 & -0.04 & 0.05 & *-0.12 \\
    Cond.3 &       &       & -0.04 & 0.06 & *-0.11 \\
    Cond.4 &       &       &       & 0.09 & -0.08 \\
    Cond.5 &       &       &       &      & **-0.17 \\
    \bottomrule
    \end{tabular}}
\end{subtable}
\hspace{12pt}
\begin{subtable}[c]{0.3\textwidth}
    \centering
    \subcaption{Medium complexity inputs}
    \resizebox{\textwidth}{!}{
    \begin{tabular}{crrrrr}
    \toprule
    & \multicolumn{1}{c}{C.2} & \multicolumn{1}{c}{C.3} & \multicolumn{1}{c}{C.4} & \multicolumn{1}{c}{C.5} & \multicolumn{1}{c}{C.6} \\
    \midrule
    Cond.1 & 0.00 & 0.00  & 0.00  & 0.07 & -0.09 \\
    Cond.2 &      & 0.00  & -0.01 & 0.06 & -0.09 \\
    Cond.3 &      &       & -0.01 & 0.06 & -0.09 \\
    Cond.4 &      &       &       & 0.07 & -0.08 \\
    Cond.5 &      &       &       &      & *-0.15 \\
    \bottomrule
    \end{tabular}}
\end{subtable}
\hspace{12pt}
\begin{subtable}[c]{0.3\textwidth}
    \centering
    \subcaption{High complexity inputs}
    \resizebox{\textwidth}{!}{
    \begin{tabular}{crrrrr}
    \toprule
    & \multicolumn{1}{c}{C.2} & \multicolumn{1}{c}{C.3} & \multicolumn{1}{c}{C.4} & \multicolumn{1}{c}{C.5} & \multicolumn{1}{c}{C.6} \\
    \midrule
    Cond.1 & -0.02 & -0.05 & -0.01 & 0.02 & **-0.13 \\
    Cond.2 &       & -0.02 & 0.01  & 0.05 & **-0.11 \\
    Cond.3 &       &       & -0.04 & 0.07 & -0.09 \\
    Cond.4 &       &       &       & 0.03 & -0.12 \\
    Cond.5 &       &       &       &      & **-0.16 \\ 
    \bottomrule
    \end{tabular}}
\end{subtable}\\
\vspace{6pt}
Using pairwise Mann-Whitney U test with Bonferroni correction: 
* $p<0.05$, ** $p<0.01$
\end{table*}

\begin{table*}[t]
\centering
\small
\caption{Differences of medians (row minus column) on the diversity metric among the six conditions, by input complexity.}
\label{tab:mean-difference-diversity-by-complexity}
\begin{subtable}[c]{0.3\textwidth}
    \centering
    \subcaption{Low complexity inputs}
    \resizebox{\textwidth}{!}{
    \begin{tabular}{crrrrr}
    \toprule
    & \multicolumn{1}{c}{C.2} & \multicolumn{1}{c}{C.3} & \multicolumn{1}{c}{C.4} & \multicolumn{1}{c}{C.5} & \multicolumn{1}{c}{C.6} \\
    \midrule
    Cond.1 & 0.02 & 0.09 & 0.10 & **-0.09 & **0.14 \\
    Cond.2 &      & 0.07 & 0.08 & **-0.11 & *0.12 \\
    Cond.3 &      &      & 0.01 & **-0.18 & 0.05 \\
    Cond.4 &      &      &      & **-0.19 & 0.04 \\
    Cond.5 &      &      &      &         & **0.23 \\
    \bottomrule
    \end{tabular}}
\end{subtable}
\hspace{12pt}
\begin{subtable}[c]{0.3\textwidth}
    \centering
    \subcaption{Medium complexity inputs}
    \resizebox{\textwidth}{!}{
    \begin{tabular}{crrrrr}
    \toprule
    & \multicolumn{1}{c}{C.2} & \multicolumn{1}{c}{C.3} & \multicolumn{1}{c}{C.4} & \multicolumn{1}{c}{C.5} & \multicolumn{1}{c}{C.6} \\
    \midrule
    Cond.1 & 0.01 & 0.04 & 0.06  & **-0.06 & 0.15 \\
    Cond.2 &      & 0.04 & 0.05  & **-0.06 & 0.14 \\
    Cond.3 &      &      & 0.01  & *-0.10  & 0.11 \\
    Cond.4 &      &      &       & **-0.11 & 0.09 \\
    Cond.5 &      &      &       &         & **0.20 \\
    \bottomrule
    \end{tabular}}
\end{subtable}
\hspace{12pt}
\begin{subtable}[c]{0.3\textwidth}
    \centering
    \subcaption{High complexity inputs}
    \resizebox{\textwidth}{!}{
    \begin{tabular}{crrrrr}
    \toprule
    & \multicolumn{1}{c}{C.2} & \multicolumn{1}{c}{C.3} & \multicolumn{1}{c}{C.4} & \multicolumn{1}{c}{C.5} & \multicolumn{1}{c}{C.6} \\
    \midrule
    Cond.1 & 0.00 & 0.04 & 0.00  & -0.07  & *0.18 \\
    Cond.2 &      & 0.04 & 0.00  & -0.07  & *0.18 \\
    Cond.3 &      &      & -0.04 & -0.11  & 0.14 \\
    Cond.4 &      &      &       & *-0.07 & 0.18 \\
    Cond.5 &      &      &       &        & **0.25 \\
    \bottomrule
    \end{tabular}}
\end{subtable}\\
\vspace{6pt}
Using pairwise Mann-Whitney U test with Bonferroni correction: 
* $p<0.05$, ** $p<0.01$
\end{table*}

\begin{figure*}[t]
\centering
\parbox[t][][t]{\figrasterwd}{
    \parbox{.52\figrasterwd}{%
        \subcaptionbox{Directly generated examples involves noises but can provide useful insights and suggestions.\label{fig:examples-direct-generation}}{\includegraphics[width=\hsize]{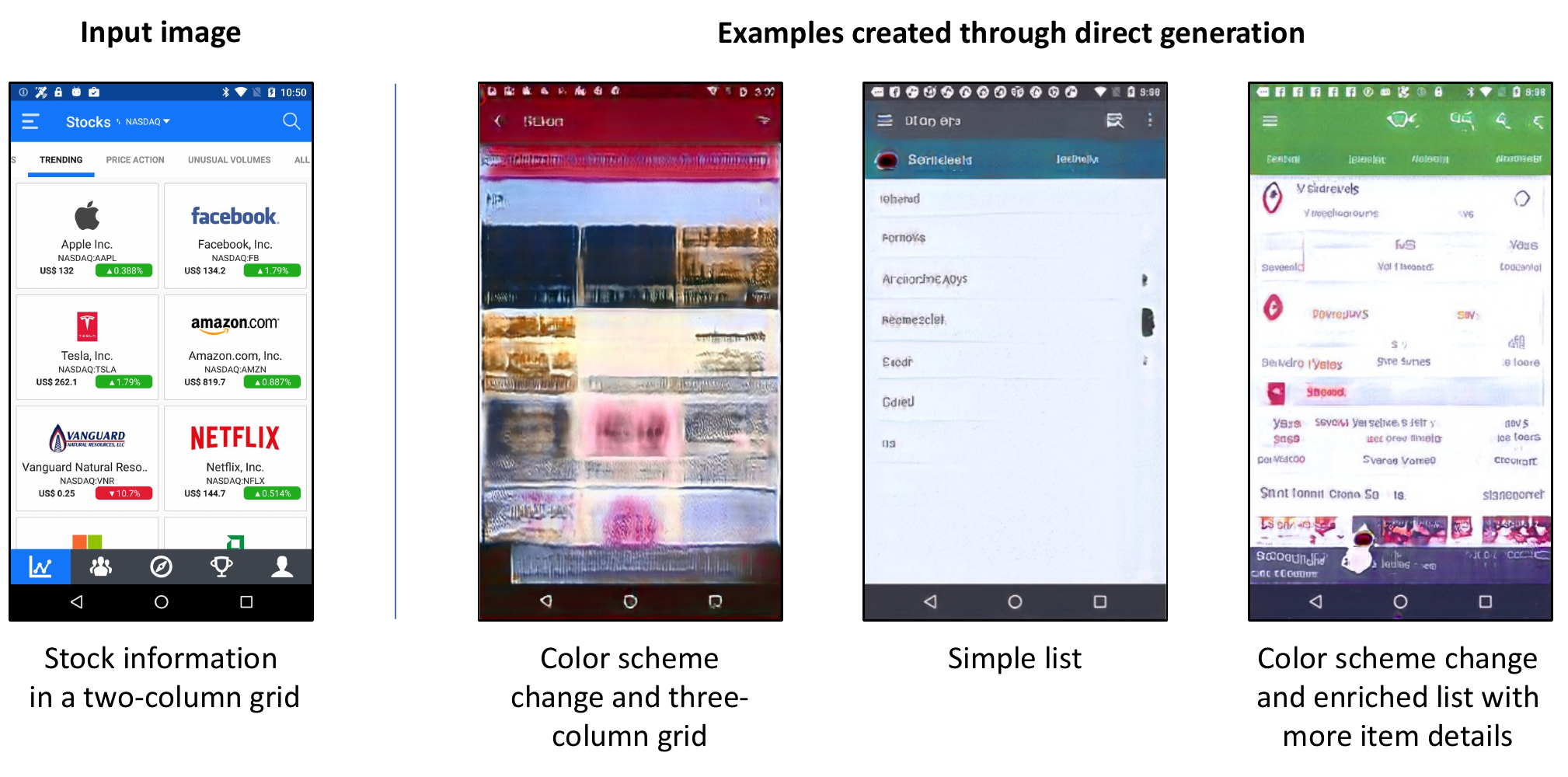}}
        \vskip1em
        \subcaptionbox{Searched examples based on generated images are cleaner and can provide useful insights and suggestions.\label{fig:examples-generation-plus-search}}{\includegraphics[width=\hsize]{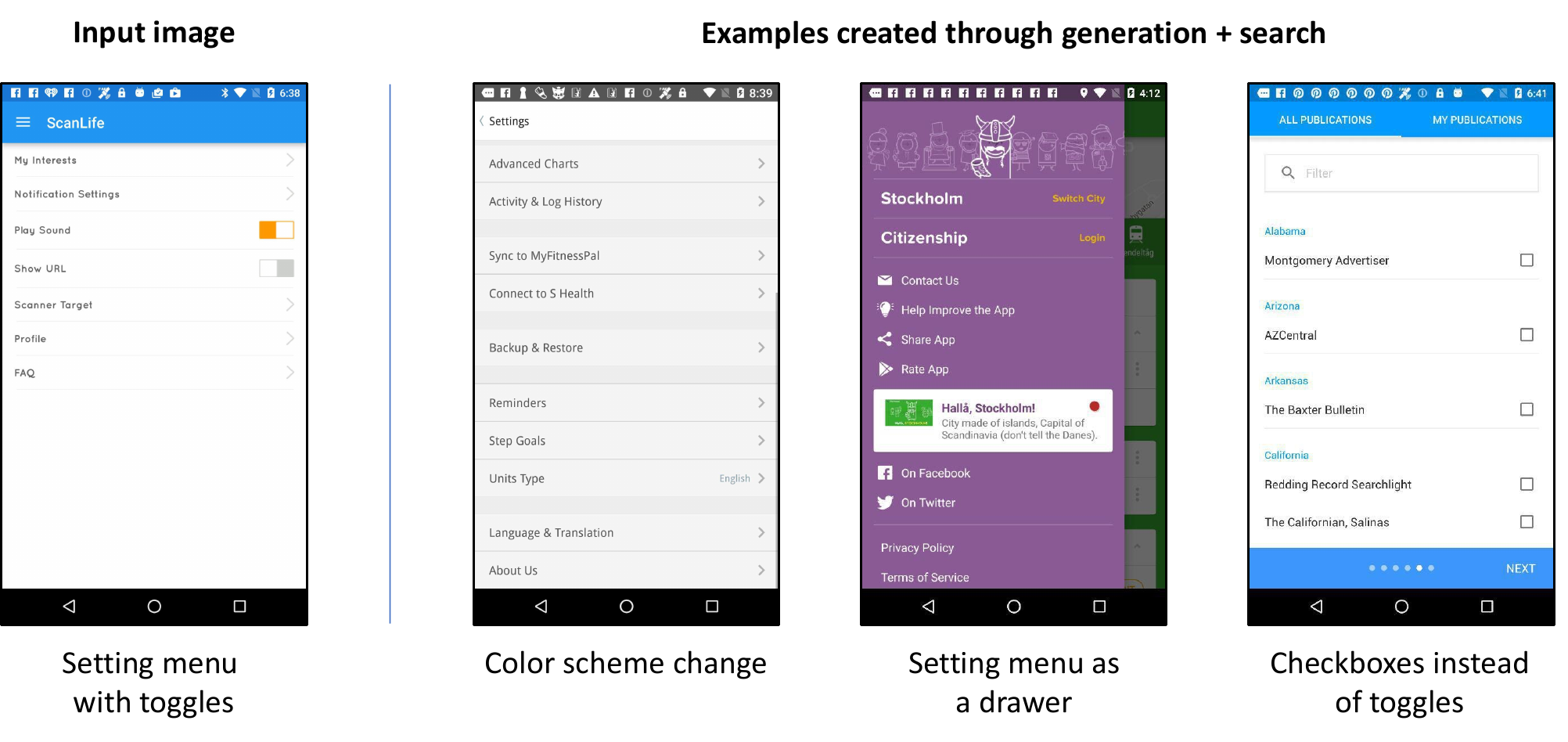}}
    }
    \hskip3em
    \parbox{.4\figrasterwd}{%
        \subcaptionbox{Search results sometimes do not reflect the intention of the style-based generation results.\label{fig:search-not-ideal}}{\includegraphics[width=\hsize]{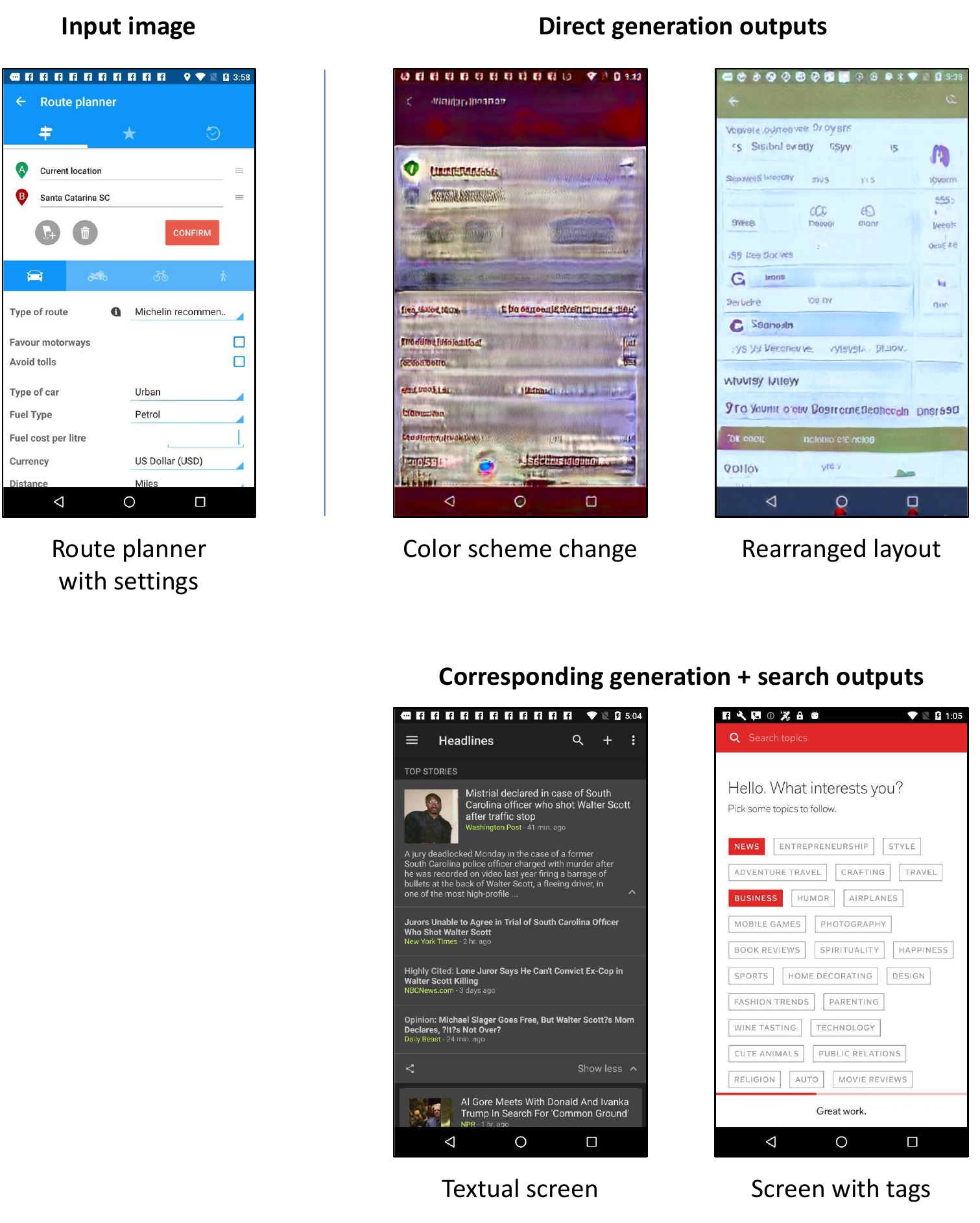}}
    }
}
\caption{Examples demonstrating insights gained from manual inspection of outputs of GANSpiration-based approaches.}
\Description{Three sets of example UI images generated through the GANSpiration-based approaches to illustrate the insights described in section 5.4.}
\label{fig:output-analysis} 
\end{figure*}

We manually inspected the outputs of the approaches used in the six experimental conditions to identify their risks and potential to support design inspiration. We found the following themes through our inspection.

\begin{itemize}
    \item Conditions 1 and 2 (i.e., generation-only): Examples generated in these two conditions resembled real UI screenshots, but contained a lot of blurry and noisy components. However, some examples generated in these conditions contained interesting variations and alternatives to the input UI. For example, Figure~\ref{fig:examples-direct-generation} shows that the directly generated examples suggested alternatives on the color scheme, layout, and item details.
    \item Conditions 3 and 4 (i.e., generation + search): Examples generated in these two conditions contained real images that addressed the noise issues in Conditions 1 and 2. Some of the returned examples also contained interesting variations to the input image (see Figure~\ref{fig:examples-generation-plus-search}). However, some of the searched images do not reflect directly the intention of the originally generated images. For example, in Figure~\ref{fig:search-not-ideal}, the generated example seems to suggest a different color scheme and a layout change, but the search resulted in somewhat irrelevant screens. 
    \item Condition 6 (i.e., search-based approach): The outputs of this condition commonly contained visually similar UI screenshots with the input, both in terms of color and layout (see Figure~\ref{fig:condition-output-comparison}, Condition 6).
\end{itemize}

\section{User Evaluation}
To understand the potential of the GANSpiration-based techniques from the perspective of professional UI/UX practitioners, we conducted a user study focusing on the practitioners' opinions on how well the results of GANSpiration may help in their design practice.

\subsection{Methods}
The user studies were conducted in June and July 2021. In this section, we describe our participants, the procedure of the study sessions, the materials used, and the analysis methods. The user study protocol is approved by the ethics committee at our institutions.

\paragraph{Participants.}
We conducted the user study with five UI/UX professionals. The participants had varying levels of experience, ranging from one to ten years as either UX researcher or UI/UX designer. They worked in different types of organizations, including two freelancers, one in a small start-up company, one in a more established medium-sized company, and one in a large multinational company. Table~\ref{tab:participants} summarizes the participants' characteristics.

\begin{table*}
\centering
\caption{Characteristics of participants in the user study}
\label{tab:participants}
\small
\begin{tabular}{cllcc}
\toprule
ID &
  \multicolumn{1}{c}{Job title} &
  \multicolumn{1}{c}{\begin{tabular}[c]{@{}c@{}}Organization\\ description\end{tabular}} &
  \begin{tabular}[c]{@{}c@{}}Years of\\ experience\end{tabular} &
  \begin{tabular}[c]{@{}c@{}}Projects\\ contributed to\end{tabular}\\
\midrule
P1 & UX researcher & Freelancer & 1 year & 1 project \\
P2 & UI/UX designer & \begin{tabular}[c]{@{}l@{}}A small-sized company developing software solutions\\ that allow the creation of printable personalized products\end{tabular} & 8 years & \textgreater 10 projects \\ %Mediaclip
P3 & UI designer & \begin{tabular}[c]{@{}l@{}}A medium-sized company developing a cloud-based\\ computer-aided design (CAD) software\end{tabular} & 3 years & 10 projects \\ %Vention
P4 & UI designer & Freelancer & 10 years & \textgreater 10 projects \\
P5 & UX researcher & \begin{tabular}[c]{@{}l@{}}A large multinational company developing business\\ management software\end{tabular} & 2 years & 2 projects\\ %Oracle
\bottomrule
\end{tabular}
\end{table*}

\paragraph{Procedure and material.}
The user studies were conducted online via the Zoom platform. Each participation took about one hour to complete. Each study session began with a short interview in which the participants were asked about their professional background and their experience of using design examples. We then presented one input UI screenshot from the sample used in the quantitative evaluation (the one that is shown in Figure~\ref{fig:condition-output-comparison}) to the participants. We told the participants to consider a scenario in which they want to get inspiration from examples to modify the design of this UI. This UI screenshot contained 11 different types of components, representing a complicated design task. After the participants familiarized themselves with this UI screenshot, we then presented the corresponding output images from all six conditions as the design examples, one after another; a sample of these materials was shown in Figure~\ref{fig:condition-output-comparison}. The participants were asked to examine each set of the design examples and to provide feedback on (1) their relevance to the design task, (2) the diversity of the design examples, (3) the effectiveness of the design examples for inspiration, and (4) general positive and negative perceptions of the examples. The order of these six conditions was randomized among the participants to mitigate the order effects; the participants were also not aware of the conditions in which the examples were obtained when examining the examples. After the participants examined all six conditions, we described the GANSpiration techniques and asked the participants about their general perception of the generated design examples.

\paragraph{Analysis.}
Two researchers watched recordings of all the study sessions and took detailed notes regarding the participants' comments and reactions. We then conducted an inductive thematic analysis~\citep{Vaismoradi2013} on the combined notes. Particularly, two researchers first independently conducted open coding to consolidate different aspects discussed by the participants. Then we held three two-hour meetings to discuss our observations and extract common themes from our codes. During this analysis, we focused on identifying the positive and negative aspects that the participants mentioned about the design examples resulting from the four categories of the experimental conditions: direct style-based generation of UI images (Conditions 1 and 2), style-based generation then search of real UI examples (Conditions 3 and 4), random selection (Condition 5), and similarity-based search (Condition 6).

\subsection{Results}
All of the participants indicated that they frequently looked for examples in their design projects. A common reason that the participants search and examine examples was to learn from the examples and borrow elements from them. For example, P3 mentioned ``Every single time when I need to design an interface, I will go to all these different reference websites. I always start from there to see what a possible solution is out there. Because it saves my time not to reinvent the wheel.'' P2 talked about the granularity in which the design examples may help, saying ``Even for a specific project, every part of the project, I need to assess what exists right now and what I can learn from what is already existing.'' P4 also emphasized the importance of examples by mentioning the efforts she often spends on it, saying ``Every project I have to search for examples for inspiration. I will spend lots of time on it.''

\subsubsection{Perceptions on the directly generated examples (Conditions~1 and 2).}
The negative aspects of these generated examples mentioned by the participants were most concerned with the quality of the images and the noises included in the UI. For example, P3 mentioned: ``It looks a bit strange to me because all the UI elements are not real UI elements. They just mimic the shapes. It doesn't provide a lot of realistic details for me to get a record for my design.'' P5 also declared, ``If there is not any noise in the images, it is a good idea to have this technique.'' 

Participants also discussed the potential of the generated examples. Particularly, these generated images provided enough abstraction and omitted unnecessary details to convey the design concept while eliciting creativity; this is similar to the effect of using low-fidelity prototypes such as sketches and wireframes. For example, P2 mentioned: ``The idea is pretty close to what I am looking for, overall, conceptually... Since they are all pretty abstract, they are already helping me more, so that I can see the shapes and stuff ... to think out of the box.'' Additionally, the collections of generated images were perceived as a gateway to gather ideas of a wider range of design styles, presented in the context of the designer's work. For example, P3 said: ``If it can generate a lot of good information online that we don't have time to look for, and it manages to make a merge of all the relevant interface, I can use it as a reference to see the trend on color palette and layouts.''

\subsubsection{Perceptions on the examples created from generation and then search (Conditions~3 and 4).}

All but one of the participants agreed the examples can be effective for inspiration. Echoing our quantitative results, participants appreciated that the examples are both diverse and relevant. For instance, P1 mentioned: ``It is diverse, we have different designs... They are also relevant since I can see some of the same patterns..'' Focusing on the diversity of the examples on the level of both the overall UI structure and the specific components, P5 said, ``These are effective for inspiration as example images. Since they include both text and big image and different kinds of components inside.'' Participants also noticed some of the components in the images could be inspiring; those components are not in the original input image but are relevant to the design task. P4 mentioned, ``Image number 14 at the top right there is an alert icon which I can use in my design and Number 3 also has a search icon.''  P3 also stated, ``There are some images which have tabs, which I can use as references for the tab of the main image, and also the toggle switch, radio buttons, ...''

The participant (P2) who thought the examples are not effective for inspiration was mainly concerned that the overall design style of the output examples is homogeneous and a bit outdated. She mentioned: ``It is all the same [Google] Material Design style and nothing exciting -- it is very similar to what I would work on already [in the input image].'' This comment pointed out the limitations of using UI screenshots from the same platform for inspiration. Some participants also voiced other concerns related to the current GANSpiration-based techniques. For example, P3 mentioned that she wished to have the examples contain more UI screenshots from the same page type and the same application domain: ``I did not find any image that included extended forms or any transportation apps.'' P1 and P4 also would like to see more recent and modern designs in the examples; this comment, along with P2's concerns, highlighted the important role of the dataset used for generating the examples and searching for existing designs.

\subsubsection{Perceptions on the randomly selected examples (Condition~5).}

All participants were impressed by the obvious diversity of the examples. For instance, P2 stated, ``I like that the color schemes are getting different. It has different layouts from different stuff.'' However, they also noticed the randomness of the examples. P3 said, ``There are a lot of things that are not relevant here. Some of the examples are definitely noises.'' P2 also said, ``I do think that those layouts, that are different, are not related to the design I am looking for.''

\subsubsection{Perceptions on the examples selected from similarity-based search (Condition~6).} 

All participants agreed that the examples are similar to the source image, but not diverse enough for inspiration. P3 said, ``The first thing that I noticed is the similar color palette and elements.'' P2 also said, ``All of them are lists and have the same colors, not very helpful but relevant.'' P4 also mentioned, ``Although the information is clear and easy to use, [the examples are] too boring.''

\vspace{9pt}
Overall, the participants considered the GANSpiration-based approaches, particularly the ones that output full-fledged UI screenshots (i.e., conditions 3 and 4), as viable ways to gain inspiration in their practical design workflows. Participants mentioned that the output examples from these approaches can help them widen their perspective and obtain immediate access to relevant and diverse inspirational sources. For example, P1 indicated ``This tool would help in the competitive analysis part. If it has more trendy images it could be really helpful.'' Some participants also commented on the ability of our approaches in helping the designers think differently and unconventionally. For example, P2 stated that ``I could use this technique... to think out of the box.''

\section{Discussion}
In both the quantitative evaluation and the qualitative user study, we found that GANSpiration-based approaches are able to generate design examples that are both diverse and relevant to the screenshot of the input design. The quantitative results indicated that our approach can generate more relevant examples than random suggestion, while providing more diverse examples than similarity-based suggestion. In the user study, our participants commented on the balance between diversity and relevancy as a desirable attribute. Participants generally preferred the GANSpiration-based approaches, particularly the ones that suggests full-fledged UI screenshots based on the generated examples, over random examples and similarity-based examples. The user evaluation also indicated that our approach is able to help designers broaden their horizons and get inspired. While the user study focused on an in-depth examination of one input image (representing a complicated design task), the quantitative evaluation covered a wider range of design complexity and indicated that our approach may work particularly well for low complexity and high complexity inputs. Overall, these results demonstrated that GANSpiration can help achieve an intricate balance between design drift and design fixation when providing examples for inspiration in the challenging context of user interface design. In the rest of this section, we discuss the implications of our study results for designing tools that leverage generative techniques such as GANSpiration for design inspiration.

\subsection{Style-based image generation provides inspiration at different granularity levels}
Our user study results indicated that the style-based generation technique adopted by GANSpiration is able to provide design inspiration on three levels: (1) the coarse level that provides ideas for layout or structural changes, (2) the middle level that suggests component design alternatives, and (3) the fine level that proposes different aesthetics such as color schemes. This is made possible by the ability of StyleGAN to alter the input image based on different granularity levels (i.e., different spatial resolutions) of the target `style images'. The three aspects of inspiration were all appreciated by the participants during the user study. The tool design that leverages GANSpiration-based approaches can explicitly incorporate these levels of design inspiration. Particularly, a design inspiration tool can indicate and explain the intention of the suggested examples by checking the granularity level of the style image used for style merge. Based on this information, a descriptive label of inspiration granularity can be assigned to each example. This way, if the designers have a particular concern when searching for inspiration (e.g., need to find a different layout but using the same color scheme), they would be able to better focus on such examples. Additionally, it is also possible to give the designers control over the granularity level of style merge.

\subsection{The visual quality of the generated image is an important factor for inspiration}
Our results revealed that the participants preferred full-fledged UI screenshots over directly generated images that typically include noises, although the direct generation and the generation-then-search conditions achieved the same level in the diversity and relevance metrics. While some participants were impressed that the directly generated images look like a UI, our results indicated that the visual quality of the generated images does affect how well the designers perceive the examples. In this study, we retrieved the most visually similar UI screenshots to the directly generated images to address this issue. However, our manual inspection revealed that the search results based on the generated images do not always match the intention of the directly generated results. To further resolve these issues, techniques for identifying components on generated images could be investigated. With such techniques, the quality of the individual components can be improved to make the generated images look more similar to real ones. Further, if components were identified, the directly generated images can be converted to wireframes in order to provide layout or structural suggestions. Color schemes of the components can also be matched to the input image to provide direct alternatives.

\subsection{A diverse and relevant training dataset would help generate more insightful examples}

Our study relies on the Rico dataset, which is created in 2017. Some of our participants voiced that the examples retrieved do not reflect the most recent design trends, thus limit the inspirational power of the examples. This result indicates that the dataset used for training the generative model, as well as the dataset used for retrieving real UI screenshots, are important factors to consider. Potential solutions to this problem include using only the newest apps in the dataset and collecting datasets from the most recent design sharing platforms (e.g., dribbble.com). Additionally, using a merged dataset including UI screenshots from different platforms (e.g., Android, iOS, desktop app, web app, etc.) with different design frameworks/systems~\cite{dshandbook} would help avoid platform or framework-specific design stereotypes. %However, the size of the resulted dataset may not be sufficient for achieving a satisfactory generation quality. 

\subsection{Combine generative models with other techniques}
In our experimental design, conditions 3 and 4 (i.e., generation + search) are our first attempts to combine the pure generative approach with other techniques to provide examples for effective inspiration. These attempts can be further enriched and expanded. Particularly, our participants seemed to desire examples from the same application domain (e.g., route planning) and/or focusing on the same type of page (e.g., configuration page) as the input UI page. Thus, it could be useful to perform the style-merging generation using the style images from the same application domain and/or page type as the input image. Moreover, combining the generative technique with textual thematic specifications that describe characteristics of the desired examples (e.g., the ones similar to~\cite{Ritchie2011}) would give designers more control over the returned examples and support a more effective inspiration. Finally, techniques that can highlight interesting areas in the examples related to the suggested alternatives to the input image would also facilitate a more efficient exploration for serendipitous and targeted inspiration.

\section{Threats to Validity and Limitations}
As mentioned before, the dataset we used in this study (i.e., the Rico dataset) is published about five years ago and only included screenshots of Android applications. Although it is a large dataset that is frequently used in studies involving UI design artifacts, we do not know the inspirational power of GANSpiration-based approaches if a newer dataset or a dataset on another platform was used. We recognize that building a large-scale UI dataset is a non-trivial task. Even with the old dataset, our study demonstrated the potential of GANSpiration in supporting both serendipitous and targeted inspirations.

Moreover, our approach treated user interface design as static images. We acknowledge that the design artifacts can be presented in many other formats (e.g., a tree of UI components~\cite{Wu2021}) and may include other information such as animations and page transitions. We chose to focus on static images because they are one of the most frequent types of inspirational sources currently used by the designers, supported by tools such as design galleries (e.g., Dribbble). Future work may investigate how similar approaches can be developed to incorporate more structured and richer representations of user interface design artifacts.

Additionally, our user study only included five participants. Moreover, while the evaluation was based on a realistic scenario, it only included one input UI and may not be able to incorporate all the real-world inspirational challenges related to UI design. The small sample size is partially due to the challenges we experienced in recruiting participants during the pandemic. However, our participants were all professional practitioners and represented diverse UI/UX-related expertise. With their professional experiences, they also reflected on their practice when examining the examples during the user study, providing a real-world perspective.

Finally, we acknowledge that UI/UX design involves complex workflows, diverse tasks, and various considerations. We do not claim that our work can possibly address all these aspects. Instead, we targeted the particularly challenging problem of design inspiration and demonstrated the potential of a creativity support approach for this context that leverages a style-based generative machine learning technique. We encourage future work to investigate the applicability of this line of techniques in other tasks and aspects of UI/UX design.
\section{Conclusion}
Our proposed GANSpiration-based approaches aim to provide concrete creativity support by balancing both targeted and serendipitous inspiration in user interface design. The evaluation studies highlighted the capacity of GANSpiration in generating examples that are both relevant to the designers' work at hand and diverse for avoiding design fixation. Professional UI/UX practitioners appreciated such techniques as viable support in their day-to-day design practice. Our results also revealed possible improvements and design implications when such a generative technique is used for supporting design inspiration. Overall, our work demonstrates the potential of applying style-based generative machine learning techniques in the challenging context of design inspiration and creativity support. It opens a new direction and paves the road for future efforts in using advanced intelligent technology for supporting the creative, but at the same time constraint, design practice.

\begin{acks}
We thank the participants for their time and valuable insights. The project is partially supported by the Natural Sciences and Engineering Research Council of Canada (NSERC) Discovery Grant Program [RGPIN-2018-04470] and Fonds de Recherche du Qu\'{e}bec – Nature et technologies (FRQNT) Team Research Project Grant [2022-PR-299099].
\end{acks}

\bibliographystyle{ACM-Reference-Format}
\bibliography{reference}

\end{document}